\documentstyle[12pt,epsfig]{article}
\let\section=\subsection     \let\subsection=\subsubsection                
\newcommand{\case}[2]{\mbox{\footnotesize $\displaystyle \frac{#1}{#2}$}}
\newcommand{\lsim}{\mathrel{\rlap{\lower3pt\hbox{\hskip0pt$\sim$}}
\raise2pt\hbox{$<$}}}
\newcommand{\gsim}{\mathrel{\rlap{\lower3pt\hbox{\hskip0pt$\sim$}}
\raise2pt\hbox{$>$}}}

\begin{document}
\begin{center}
   {\large \bf TEMPERATURE, CHEMICAL POTENTIAL \\
        AND THE $\rho$--MESON
}\\[2mm]
 C.D.~ROBERTS and S.M.~SCHMIDT\\[5mm]
   {\small \it Physics Division, Argonne National Laboratory,\\
Argonne IL 60439-4843, USA   \\[8mm] }
\end{center}

\addtocounter{section}{1}

\hspace*{-\parindent}{\bf
\arabic{section}.~Introduction.}\hspace*{0.5\parindent} Models of QCD must
confront nonperturbative phenomena such as confinement, dynamical chiral
symmetry breaking (DCSB) and the formation of bound states.  In addition, a
unified approach should describe the deconfinement and chiral symmetry
restoring phase transition exhibited by strongly-interacting matter under
extreme conditions of temperature and density.  Nonperturbative
Dyson-Schwinger equation (DSE) models~\cite{cdragw,echaya} provide insight
into a wide range of zero temperature hadronic phenomena; e.g., non-hadronic
electroweak interactions of light- and heavy-mesons~\cite{misha}, and diverse
meson-meson~\cite{tandy} and meson-nucleon~\cite{bloch1} form factors.  This
is the foundation for their application at
nonzero-($T,\mu$)~\cite{echaya},\cite{bender1}-\cite{bloch}.  Herein we
describe the calculation of the deconfinement and chiral symmetry restoring
phase boundary, and the medium dependence of $\rho$-meson properties.  We
also introduce an extension to describe the time-evolution in the plasma of
the quark's scalar and vector self energies based on a Vlasov equation.

\medskip\addtocounter{section}{1}

\hspace*{-\parindent}{\bf \arabic{section}.~Dyson-Schwinger Equation at
Nonzero-($T,\mu$).}\hspace*{0.5\parindent} The dressed-quark DSE at
nonzero-($T,\mu$) is
\begin{eqnarray}\label{qprop}
S^{-1}(\tilde p_k) = i\vec{\gamma}\cdot\vec{p}\, A(\tilde p_k) 
+ i\gamma_4 \omega_{k_+}\, C(\tilde p_k) 
+ B(\tilde p_k)=i\vec{\gamma}\cdot\vec{p}\,+ i\gamma_4
\omega_{k_+}\,+\Sigma(\tilde p_k) \,,
\end{eqnarray}
where $\tilde p_k = (\vec{p},\omega_{k_+})$, $\omega_{k_+}= \omega_k + i
\mu$, and $\omega_{k} = (2k + 1)\pi T$ is the quark's Matsubara frequency.
The complex-valued scalar functions: $A(\vec{p},\omega_{k_+})$,
$B(\vec{p},\omega_{k_+})$ and $C(\vec{p},\omega_{k_+})$, depend only on
$(|\vec{p}|^2,\omega_{k_+}^2)$.  With a given dressed-gluon propagator the
solutions are determined by
\begin{eqnarray}
\label{Bfunc}
B(\tilde p_k)-m_0 &=& \frac{8}{3}\int\frac{ d^4 \vec{q}}{(2\pi)^4} D(\tilde
p_k-\tilde q_k)\frac{B(\tilde q_k)}{\tilde q_k^2 \,C^2(\tilde q_k)+B^2(\tilde
q_k)},\\\label{Afunc}
(C(\tilde p_k)-1)\tilde p_k^2&=&\frac{4}{3}\int\frac{ d^4 \vec{q}}{(2\pi)^4}
D(\tilde p_k-\tilde q_k)\frac{\tilde p_k\cdot \tilde q_k C(\tilde
q_k)}{\tilde q_k^2 \,C^2(\tilde q_k)+B^2(\tilde q_k)}\,\,,
\end{eqnarray}
where herein we only consider models where $A(p)=C(p)$.  It is the interplay
between the functions $B$ and $C$ that leads to confinement, realised via the
absence of a Lehmann representation for the dressed-quark $2$-point
function~\cite{cdragw,echaya},\cite{bender1}-\cite{maris}.  $B\neq 0$ in the
chiral limit signals DCSB.

To provide an illustrative solution of the quark DSE we employ an Ansatz for
the scalar function characterising the dressed-gluon
propagator~\cite{thermo}:
\begin{equation}\label{Gpro}
D(p) = 3\pi^2\frac{\eta^2}{T}\,\,\delta_{k0}\,\delta^3(p)\,.
\end{equation}
The infrared enhancement in this choice ensures quark confinement and DCSB.
As an infrared dominant model, Eq.~(\ref{Gpro}) does not represent well the
interaction away from $(\tilde p_k - \tilde q_k)^2\simeq 0$ and that
introduces some model-dependent artefacts.  However, they are easily
identified and the model yields qualitatively reliable results, preserving
features of more sophisticated studies.

\begin{figure}[t]
\centering{\
\epsfig{figure=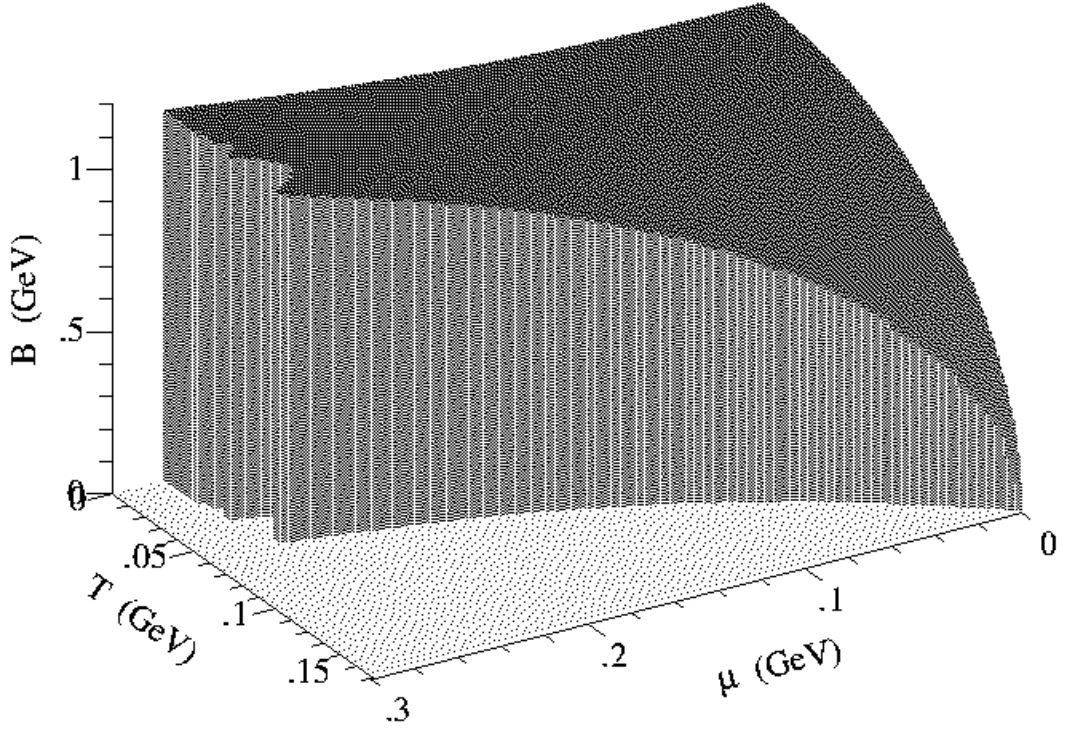,width=13.0cm,angle=-90}}\vspace*{-3.0\baselineskip}

\caption{Chiral order parameter: $B(\tilde p_k)$ obtained from
Eq.~(\protect\ref{Bfunc}), as a function of $(T,\mu)$: ($T_c=158$
MeV,$\mu=0$), ($T=0,\mu_c=275$ MeV).  As described in the text, the phase
boundary is fixed by the condition ${\cal B}(T,\mu)= 0$.\label{fig1}}
\end{figure}

Using Eq.~(\ref{Gpro}) in Eqs.~(\ref{Bfunc}-\ref{Afunc}) we obtain a system
with two phases.  The Nambu-Goldstone (NG) phase is characterised by
dynamically broken chiral symmetry and confined dressed-quarks. The
alternative Wigner-Weyl (WW) solution describes a phase of the model with
restored chiral symmetry and deconfinement.  In studying the phase transition
one must consider the relative stability of the confined and deconfined
phases, which is measured by the $(T,\mu)$-dependent pressure difference
between the two distinct phases: ${\cal B}(T,\mu) = P[S_{\rm NG}] - P[S_{\rm
WW}]$. ${\cal B}(T,\mu) >0$ indicates the stability of the confined
(Nambu-Goldstone) phase and hence the phase boundary is specified by that
curve in the $(T,\mu)$-plane for which ${\cal B}(T,\mu)= 0\,$.  The critical
line is depicted in Fig.~\ref{fig1}. The phase transition is first order for
any non-zero $\mu$ and second order for $\mu=0$. The model has mean field
critical exponents, which is a feature of the rainbow-ladder
truncation~\cite{arnea}.  The study of thermodynamic properties shows that it
is essential to keep scalar and vector self-energies as well as their
momentum dependence~\cite{bloch,thermo}.

Mesons are quark-antiquark bound states and their masses are obtained by
solving the Bethe-Salpeter equation~\cite{pieter}.  Here we focus on the
vector channel and employing Eq.~(\ref{Gpro}) the eigenvalue equation for the
bound state mass is~\cite{basti}
\begin{equation}
\label{rhomass}
\frac{\eta^2}{2}\,{\sf Re}\left\{ \sigma_B(\omega_{0+}^2
        - \case{1}{4} M_{\rho\pm}^2)^2 
- \left[ \pm \,\omega_{0+}^2 - \case{1}{4} M_{\rho\pm}^2\right]
        \sigma_C(\omega_{0+}^2- \case{1}{4} M_{\rho\pm}^2)^2 \right\}
= 1\,,
\end{equation}
where 
$\sigma_{B,C}(\tilde p_k^2)= \{B(\tilde p_k^2),C(\tilde p_k^2)\}/[\tilde
p_k^2\,C^2(\tilde p_k^2) + B^2(\tilde p_k^2)].
$
The equation for the $\rho$-meson's transverse component is obtained with $[-
\omega_{0+}^2 - \case{1}{4} M_{\rho-}^2]$ in Eq.~(\ref{rhomass}) and in the
chiral-limit yields $ M_{\rho-}^2 = \case{1}{2}\,\eta^2,\;\mbox{{\it
independent} of $T$ and $\mu$.}$ This is the $T=0=\mu$ result of
Ref.~\cite{mn83}.  Even for nonzero current-quark mass, $M_{\rho-}$ changes
by less than 1\% as $T$ and $\mu$ are increased from zero toward their
critical values.  Its insensitivity is consistent with the absence of a
constant mass-shift in the transverse polarization tensor for a gauge-boson.
For the longitudinal component one obtains in the chiral limit:
\begin{equation}
M_{\rho+}^2 = \case{1}{2} \eta^2 - 4 (\mu^2 - \pi^2 T^2)\,.
\end{equation}
The results for the medium-dependence of the $\rho$ meson are summarised in
Fig.~2. As in the case of the dressed-quark mass function, the response to
increasing $T$ and $\mu$ is anti-correlated: the $\rho$- mass decreases with
increasing chemical potential and increases with temperature. This
anti-correlation leads to an edge along which the $T$ and $\mu$ effects
compensate and the mass remains unchanged up to the transition point.

\begin{figure}[t]
\centering{\
\epsfig{figure=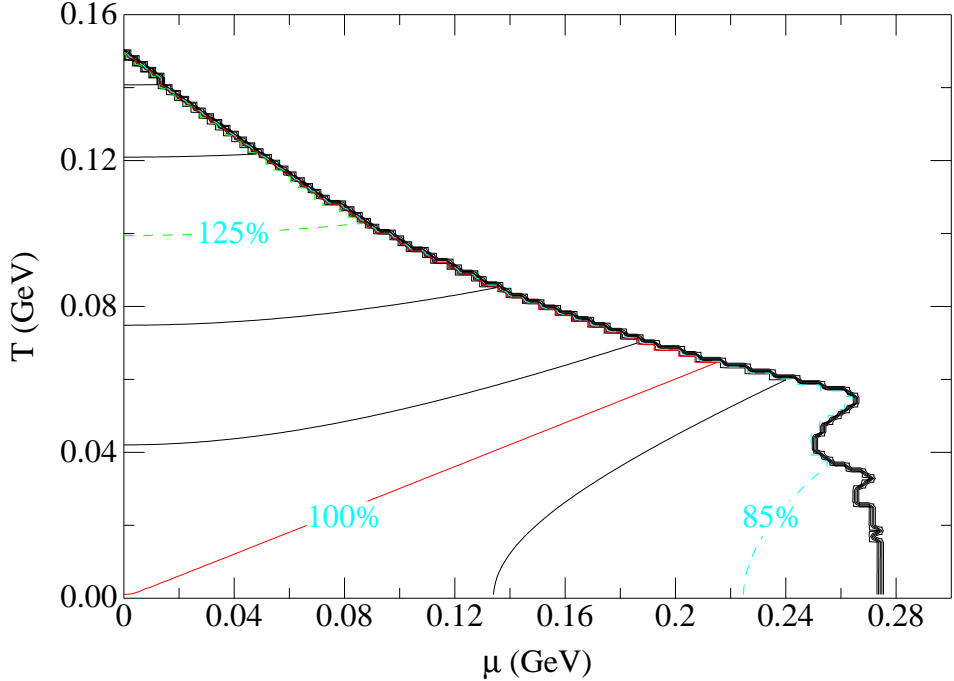,width=11.0cm,angle=-90}}\vspace*{-\baselineskip} 

\caption{$M_\rho$ as a function of $(T,\mu)$.\label{fig2}}
\end{figure}

\medskip\addtocounter{section}{1}

\hspace*{-\parindent}{\bf \arabic{section}.~Nonequilibrium
Application.}\hspace*{0.5\parindent} The time evolution of the self energies
can be studied using Vlasov's equation
\begin{equation}\label{vlasov}
\partial_t\, f(p,x) +\partial_p\, E(p,x)
\partial_x\, f(p,x) - \partial_x\, E(p,x)
\partial_p\, f(p,x)=0\,\,.
\end{equation}
Solving this equation is complicated for two reasons.  (i) The energy is a
functional of the scalar and vector self energies, which in general are
nonzero and momentum-dependent.  While the scalar self energy is small in the
plasma phase due to chiral symmetry restoration, the vector self energy
remains significant~\cite{thermo}.  (ii) The absence of a Lehmann
representation for the dressed-quark propagator in the confined phase
precludes the existence of a single particle distribution function, $f$, in
this phase.  Therefore a conventional kinetic theory is only reasonable in
the deconfined phase.  This situation is adequately represented in DSE
models; e.g., Refs.~\cite{bender1} describe a quark's $(T,\mu)$-evolution
from a confined to a propagating mode, and Ref.~\cite{thermo} makes use of
this evolved quasiparticle behaviour in calculating the plasma's
thermodynamic properties.  Therefore, approaching the phase boundary from the
plasma domain we anticipate a discontinuous disappearance of the quark
distribution function, $f$.

\begin{figure}[t]
\centering{\
\epsfig{figure=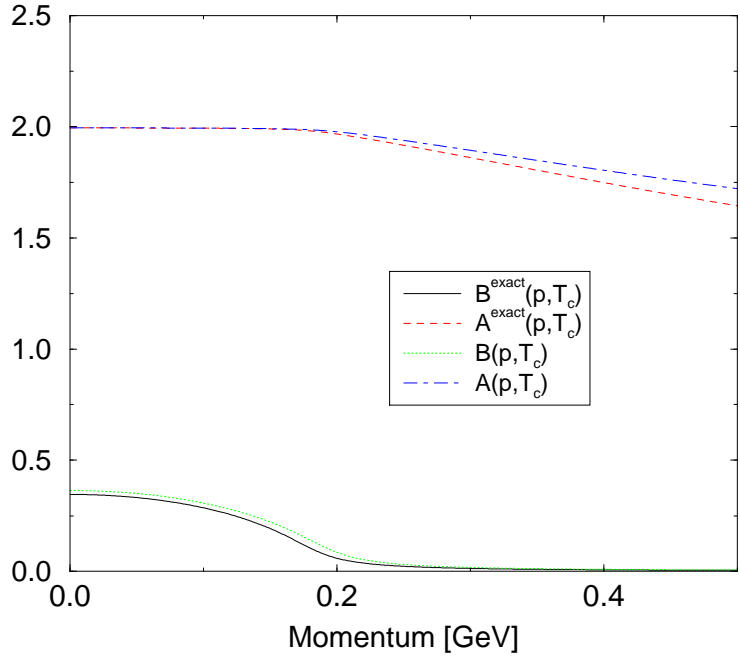,width=8.9cm,angle=-90}}\vspace*{-0.5\baselineskip} 

\caption{Momentum dependence of the quark scalar and vector self energies
obtained in the model of Eq.~(\protect\ref{Gpro}) compared with those in the
model of Eq.~(\protect\ref{IMN}).\label{fig3}}
\end{figure}

As an illustration we employ an instantaneous interaction of the form
\begin{equation}\label{IMN}
D(p) = 3\pi^2\,\eta\,\delta^3(p)
\end{equation}
to represent dynamics in the deconfined phase.  In this case the Matsubara
sum in Eq.~(\ref{Bfunc}) can be performed analytically and we obtain:
\begin{eqnarray}\label{sigS}
\Sigma^B(p,x)& =& \eta\frac{\Sigma^B(p,x) +
m_0}{(1+\Sigma^C(p,x))E^*(p,x)}[1-2f(p,x)]\,,\\
\label{sigV}
\Sigma^C(p,x) &=& \eta\frac{1}{(1+\Sigma^C(p,x))E^*(p,x)}[1-2f(p,x)]\,,
\end{eqnarray}
with the quasi particle energy: $ E^*(p,x)=\sqrt{({\vec p}^*)^2 + M^*(p,x)^2}$, the
renormalized momentum: ${\vec p}^*= {\vec p}\,(1+\Sigma^C(p,x))$, and mass:
$M^*(p,x)=m_0 + \Sigma^B(p,x)$.  As a test whether this simplification still
yields necessary and qualitatively important features, such as $C\neq 1,
B\neq m_0$, in Fig.~\ref{fig3} we compare the momentum dependence obtained in
the models specified by Eqs.~(\ref{Gpro},\ref{IMN}) in the vicinity of $T_c$.
Both functions are well reproduced and hence Eq.~(\ref{IMN}) can be used to
model the persistence of non-perturbative effects in the deconfined domain.
The solution of Eqs.~(\ref{vlasov},\ref{sigS}-\ref{sigV}) provide the
time-evolution of the quark self-energy and distribution function.

As in the case of thermal equilibrium, the vector self energy plays an
important role. Neglecting $\Sigma^C$ and the momentum dependence of
$\Sigma^B$ a simpler equation is obtained
\begin{equation}\label{vlasovnjl}
\partial_t\, f(p,x) +\frac{{\vec p}}{E(p,x)}\partial_x\, f(p,x) - m(x)
\partial_x m(x)\partial_p\, f(p,x)=0,
\end{equation}
with $m(x)$ the quark mass obtained as a solution of the gap equation in
models without confinement.  This equation has been widely studied;
e.g. Refs.~\cite{huefner}.  However, we anticipate that the numerical
solution of Eq.~(\ref{vlasov}) will yield significantly different results
because of the presence and persistence of the vector self energy in the
deconfined domain.

\hspace*{-\parindent}{\bf Acknowledgments.}\hspace*{0.5\parindent} This
work was supported by the US Department of Energy, Nuclear Physics Division,
under contract no. W-31-109-ENG-38, the National Science Foundation under
grant nos. INT-9603385 and PHY97-22429, and benefited from the resources of
the National Energy Research Scientific Computing Center.
S.M.S. acknowledges financial support from the A. v. Humboldt foundation.

\vspace*{-0.5\baselineskip}

\end{document}